\documentclass[aps,twocolumn,prd,showpacs,showkeys,preprintnumbers,superscriptaddress,nobibnotes,floatfix,longbibliography,nofootinbib]{revtex4-2}
\pdfoutput=1
\usepackage{amsmath}
\usepackage{amsfonts}
\usepackage{amssymb}
\usepackage{mathrsfs}
\usepackage{graphicx}
\usepackage{color}
\usepackage{hyperref}
\usepackage{multirow}
\usepackage{slashed}
\usepackage{epsf}

\usepackage{graphicx}
\usepackage{caption}
\usepackage{subcaption}

\usepackage{color}

\def\to{\rightarrow}

\def\tl{\tilde l}

\def\tl{\tilde l}

\def\tl{\tilde}

\begin{document}

\title{
The Light Neutralino Dark Matter in the Generalized Minimal Supergravity (GmSUGRA)
}

\author{Imtiaz Khan}
\email{ikhanphys1993@gmail.com}
\affiliation{Department of Physics, Zhejiang Normal University, Jinhua, Zhejiang 321004, China}
\affiliation{Zhejiang Institute of Photoelectronics, Jinhua, Zhejiang 321004, China}

\author{Waqas Ahmed}
\email{waqasmit@itp.ac.cn}
\affiliation{Center for Fundamental Physics and School of Mathematics and Physics, Hubei Polytechnic University}

\author{Tianjun Li}
\email{tli@mail.itp.ac.cn}
\affiliation{CAS Key Laboratory of Theoretical Physics, Institute of Theoretical Physics, Chinese Academy of Sciences, Beijing 100190, China}
\affiliation{School of Physical Sciences, University of Chinese Academy of Sciences, No. 19A Yuquan Road, Beijing 100049, China}
\affiliation{School of Physics, Henan Normal University, Xinxiang 453007, P. R. China}

\author{Shabbar Raza}
\email{shabbar.raza@fuuast.edu.pk}
\affiliation{Department of Physics, Federal Urdu University of Arts, Science and Technology, Karachi 75300, Pakistan}

\author{Ali Muhammad}
\email{alimuhammad@phys.qau.edu.pk}
\affiliation{CAS Key Laboratory of Theoretical Physics, Institute of Theoretical Physics, Chinese Academy of Sciences, Beijing 100190, China}
\affiliation{School of Physical Sciences, University of Chinese Academy of Sciences, No. 19A Yuquan Road, Beijing 100049, China}

\begin{abstract}
We investigate both the $Z$ and $H$ poles solutions for the Higgsino mass parameter $\mu>0$ and $\mu<0$ for the neutralino dark matter in light of the LHC supersymmetry searches and the direct detection dark matter experiments, LUX-ZEPLIN (LZ), in the Generalized Minimal Supergravity (GmSUGRA). Our study indicates that the latest experimental constraints from the LHC and LZ Collaborations exclude the light Higgsinos in the $Z$ and $H$ pole regions for the $\mu>0$ case. Interestingly, for the $\mu < 0$ case, a very light Higgsinos can still be consistent with the current constraints from the electroweakino searches and LZ experiment in the $Z$ and $H$ poles. Consequently, the $\mu < 0$ case appears more promising and thus requires the dedicated efforts to make definitive conclusions about their current status from the experimental Collaborations. In this framework, our findings indicate a deviation of up to $2\sigma$ from the central
value of \( a_\mu \equiv (g-2)_\mu/2 \), resonating with the experimental results reported by CMD and BDM.

\end{abstract}
\maketitle


\textbf{Introduction:}
\label{intro}
The Supersymmetric Standard Models (SSMs) are among the leading candidates for the new physics beyond the Standard Model (BSM). Despite the lack of concrete experimental evidence to date, the SSMs offer several compelling theoretical advantages. For instance,  the gauge hierarchy problem can be solved, gauge coupling unification can be achieved~\cite{gaugeunification,Georgi:1974sy,Pati:1974yy,Mohapatra:1974hk,Fritzsch:1974nn,Georgi:1974my}, and  the lightest supersymmetric particle (LSP) such as neutralino serves as a viable candidate for dark matter (DM) if R-parity is conserved~\cite{neutralinodarkmatter,darkmatterreviews}. This framework also allows for the radiative breaking of electroweak (EW) gauge symmetry due to the large top quark Yukawa coupling. Furthermore, the Minimal Supersymmetric Standard Model (MSSM) predicts the mass of the lightest CP-even Higgs boson to be within the range of 100 to 135 GeV~\cite{Slavich:2020zjv}. Consequently, the pursuit of supersymmetry (SUSY) remains a primary objective at the Large Hadron Collider (LHC), as it bridges the gap between high-energy fundamental physics and low-energy phenomenology, positioning supersymmetry (SUSY) as a leading candidate for new physics beyond the Standard Model (SM).

Following the Run-2 LHC, no conclusive evidence of SUSY has been observed, and extensive searches have imposed stringent constraints on the SSMs spectra. The investigations thus far have set the lower mass edge on stop, sbottom, first-two generation squarks, and gluino at approximately 1.25 TeV, 1.5 TeV, 2 TeV, and 2.2 TeV, respectively~\cite{ATLAS-SUSY-Search, Aad:2020sgw, Aad:2019pfy, CMS-SUSY-Search-I, CMS-SUSY-Search-II}. Consequently, the colored sparticles (superpartners of color particles), at least, must have considerable mass, around a few TeV. Despite the limitations imposed by the constraints on color sparticle masses, the existing data continue to permit the existence of models featuring an EW-scale bino-dominant LSP alongside relatively lighter sleptons, despite the heavier color particles. The SUSY models, influenced by the BSM theories, suggest that the masses of light sleptons span from a few hundred GeV to the TeV scale, as demonstrated in \cite{Ahmed:2022ude, Zhang:2023jcf, Khan:2023ryc}. The scenario involving a light neutralino, $m_{\tilde{\chi}^0_1} \lesssim m_h/2$, is particularly intriguing as it presents the possibility of the neutralino being a cold dark DM candidate. This particular scenario is especially intriguing because the SM Higgs boson can kinematically decay invisibly via $h \rightarrow \tilde{\chi}^0_1 \tilde{\chi}^0_1$. This process provides an additional DM signature within the Higgs sector. Extensive research has investigated the potential of light neutralino DM in the Phenomenological MSSM (pMSSM) and Constrained MSSM (cMSSM) and, taking into account different experimental constraints available back then\cite{Griest:1987qv,Djouadi:1996mj,Hooper:2002nq,Calibbi:2013poa,Han:2014nba,Hamaguchi:2015rxa,Cao:2015efs,Pozzo:2018anw,Wang:2020dtb,VanBeekveld:2021tgn}. Recent findings from collider experiments, such as those conducted by the ATLAS and CMS Collaborations, have provided significant insights.  These include searches for heavy Higgs bosons~\cite{ATLAS:2020zms},  investigations of charginos and neutralinos \cite{CMS:2020bfa,ATLAS:2021moa,ATLAS:2021yqv,CMS:2022sfi}, and studies on the invisible decay of the SM Higgs boson \cite{ATLAS:2022yvh}. Additionally, the Collaborations such as PICO-60, LUX-ZEPLIN (LZ), XENON-1T, and PandaX-4T have given the strong bounds on the direct detection (DD) for the DM both spin-independent (SI) and spin-dependent (SD)  scattering cross-sections \cite{XENON:2018voc,XENON:2019rxp,PICO:2019vsc,PandaX-4T:2021bab,LZ:2022lsv,PandaX:2022xas}. Given the latest and most stringent results for the spin-independent (SI) direct detection (DD) cross-sections from the LZ collaboration \cite{LZ:2022lsv}, it is imperative to reassess the parameter space of the MSSM. This is particularly relevant when considering light neutralino DM, as it has the potential to contribute to the Higgs boson invisible decay. A parameter space explored in \cite{Barman:2022jdg} while considered the frame work of pMSSM, which provides significant freedom in free parameters. In contrast, our study considers the Generalized minimal Supergravity (GmSUGRA) inspired from the supersymmetric Grand Unified Theories (GUTs), which provide a more unified and constrained theoretical framework compared to the pMSSM.  

 In this paper, we shall study the light neutralino DM within the context of the MSSM in details, taking into account both $\mu>0$ and $\mu<0$, the Higgsino mass parameter from the GmSUGRA. It is noteworthy that $\mu>0$ is typically required to explain the observed discrepancy in the muon’s anomalous magnetic moment $a_\mu \equiv (g - 2)_\mu/2$
 between experimental measurements and the SM prediction \cite{Muong-2:2023cdq, Ahmed:2021htr}. Because of the ambiguities in SM prediction, while forecasting the hadronic contribution, we stay agnostic about the sign of $\mu$. A comprehensive review of this scenario is appropriate, as the LHC gears up for Run-3 and the collection of new data. Our study aims to find interesting parameter space regions for the LHC searches during Run-3. In the  MSSM with the GmSUGRA, we concentrate on the light neutralino sector. We particularly focus on scenarios where the mass of the lightest neutralino, represented as $m_{{\tilde \chi}^0_1}$, is equal to or less than half of the SM Higgs boson mass ($m_h/2$) and half of the $Z$ boson mass ($m_z/2$). This allows for its potential contribution to the Higgs boson and the Z boson invisible decay mode. 
 We find that the latest experimental constraints from the LHC and LZ Collaborations exclude the light Higgsinos in the $Z$ and $H$ pole regions for the $\mu>0$ case. Interestingly, for the $\mu < 0$ case, a very light Higgsinos can still be consistent with the current constraints from the electroweakino searches and LZ experiment in the $Z$ and $H$ poles. Consequently, the $\mu < 0$ case appears more promising and thus requires the dedicated efforts to make definitive conclusions about their current status from the experimental Collaborations.

\textbf{GmSUGRA Model:}
\label{model}
In the GmSUGRA framework~\cite{Li:2010xr, Balazs:2010ha}, the Electroweak Supersymmetry (EWSUSY) is realized. As detailed in~\cite{Cheng:2012np,Cheng:2013hna, Li:2014dna}, this model accommodates sleptons and electroweakinos (neutralinos and charginos) with masses typically within one TeV, while squarks and/or gluinos can have masses in the several TeV range~\cite{Li:2014dna,Cheng:2012np}. In addition, at the Grand Unified Theory (GUT) scale, the GmSUGRA establishes the gauge coupling and gaugino mass relations\cite{Li:2010xr, Balazs:2010ha}, which are given as follows:
\begin{equation}
 \frac{1}{\alpha_2}-\frac{1}{\alpha_3} =
 k~\left(\frac{1}{\alpha_1} - \frac{1}{\alpha_3}\right)~,
\end{equation}
\begin{equation}
 \frac{M_2}{\alpha_2}-\frac{M_3}{\alpha_3} =
 k~\left(\frac{M_1}{\alpha_1} - \frac{M_3}{\alpha_3}\right)~.
\end{equation}
In our simplified GmSUGRA model, the index $k$ in these relations is set to 5/3. Furthermore, for simplicity, we assume that at the GUT scale ($\alpha_1=\alpha_2=\alpha_3$), the gaugino mass relation becomes
\begin{equation}
 M_2-M_3 = \frac{5}{3}~(M_1-M_3)~.
\label{M3a}
\end{equation}
Thus, the universal gaugino mass relation $M_1 = M_2 = M_3$ found in the mSUGRA is  a particular example of the GmSUGRA. Instead of three independent gaugino masses, we have two in this case. Therefore, Eq.~(\ref{M3a}) expresses $M_3$ in terms of the input parameters $M_1$ and $M_2$ as follows:
\begin{eqnarray}
M_3=\frac{5}{2}~M_1-\frac{3}{2}~M_2~.
\label{M3}
\end{eqnarray}
In Ref.~\cite{Balazs:2010ha}, the general scalar masses in the GmSUGRA have been studied in details. Here, the slepton masses are considered as independent parameters, and the squark masses derived from the $SU(5)$ GUT with an adjoint Higgs field are
\begin{eqnarray}
m_{\tl{Q}_i}^2 &=& \frac{5}{6} (m_0^{U})^2 +  \frac{1}{6} m_{\tl{E}_i^c}^2~,\\
m_{\tl{U}_i^c}^2 &=& \frac{5}{3}(m_0^{U})^2 -\frac{2}{3} m_{\tl{E}_i^c}^2~,\\
m_{\tl{D}_i^c}^2 &=& \frac{5}{3}(m_0^{U})^2 -\frac{2}{3} m_{\tl{L}_i}^2~.
\label{squarks_masses}
\end{eqnarray}
Within this context, the left-handed squark doublet masses, right-handed up-type and down-type squarks masses, and the left-handed and right-handed sleptons masses are represented by the symbols $m_{\tl Q}$, $m_{\tl U^c}$, $m_{\tl D^c}$, $m_{\tl L}$, and $m_{\tl E^c}$, respectively. Like the mSUGRA, the parameter $m_0^U$ denotes the universal scalar mass. For the light sleptons scenario, the masses of $m_{\tl L}$ and $m_{\tl E^c}$ are both constrained to be within 1 TeV. Particularly, under the assumption of $m_0^U \gg m_{\tl L/\tl E^c}$, the approximate relations for squark masses can be expressed as follows: $2 m_{\tl Q}^2 \sim m_{\tl U^c}^2 \sim m_{\tl D^c}^2$. Furthermore, under the GmSUGRA framework, the trilinear soft terms $A_U$, $A_D$, and $A_E$, as well as the Higgs soft masses $m_{\tl H_u}$ and $m_{\tl H_d}$, are all regarded as free parameters.~\cite{Cheng:2012np,Balazs:2010ha}.

\textbf{Procedure of scanning, GUT-scale parameters range, and phenomenological Constraints:} In the scanning, the ISAJET 7.85 package was employed, and random scanning of the below listed free parameters was performed. The renormalization group equations (RGEs) of the MSSM are employed in the $\overline{DR}$ regularization scheme in this package. Moreover, third-generation Yukawa couplings evolved from the weak scale to the GUT scale $M_{\rm GUT}$. It is noteworthy that  we are not considering the $g_1=g_2=g_3$ unification condition strictly at $M_{\rm GUT}$, where $g_{1}$, $g_{2}$, and $g_{3}$ are the couplings of $U(1)_{Y}$, $SU(2)_{L}$, and $SU(3)_{C}$ gauge group respectively. A few percent discrepancies from unification are explained by this flexibility and are related to the unknown GUT-scale threshold corrections~\cite{Hisano:1992jj}. All parameters, which include gauge couplings, Yukawa couplings, and Soft Supersymmetry Breaking (SSB) parameters, are then back evolve to weak scale $M_{\rm Z}$ using the boundary conditions that are specified at GUT scale. $M_{\rm U}$ is the scale at which $g_1=g_2$ in ISAJET, which uses two-loop MSSM RGEs. See \cite{ISAJET} for more specific details. Within the following parameter ranges, we perform a random scan using the parameters covered in the GmSUGRA.
\begin{eqnarray}
100 \, \rm{GeV} \leq  m_{\tilde E^c}  \leq 900 \, \rm{GeV} ~,~\nonumber \\
100 \, \rm{GeV} \leq  m_{\tilde L}  \leq 1000 \, \rm{GeV} ~,~\nonumber \\
100 \, \rm{GeV} \leq  m_0^{U}  \leq 5000 \, \rm{GeV}  ~,~\nonumber \\
80 \, \rm{GeV} \leq  M_1  \leq 1000 \, \rm{GeV} ~,~\nonumber \\
100\, \rm{GeV} \leq  M_2   \leq 1500 \, \rm{GeV} ~,~\nonumber \\
100 \, \rm{GeV} \leq  m_{\tilde H_{u,d}} \leq 5000 \, \rm{GeV} ~,~\nonumber \\
-1000 \, \rm{GeV} \leq  A_{E} \leq 1000 \, \rm{GeV} ~,~\nonumber \\
-6000 \, \rm{GeV} \leq  A_{U}=A_{D} \leq 5000 \, \rm{GeV} ~,~\nonumber \\
2\leq  \tan\beta  \leq 60~.~
 \label{input_param_range}
\end{eqnarray}
We further choose $m_t = 173.3 \, \mathrm{GeV}$ \cite{:2009ec}, and take into account both positive ($\mu > 0$) and negative ($\mu < 0$) values of $\mu$. Note, our findings are not very sensitive to one or two sigma variations in $m_t$ value \cite{bartol2}. In the following, we designate the universal $A_{U}$, $A_D$, and $A_E$. Using the Metropolis-Hastings algorithm described in \cite{Belanger:2009ti}, we explore parameter space. The data point have been collected in agreement with the Radiative Electroweak Symmetry Breaking (REWSB) as well as the lightest neutralino being the lightest Supersymmetric Particle (LSP).

Additionally, we incorporate the constraints on the charged sparticle masses set from the LEP2 experiments ($\gtrsim 100$ GeV) \cite{Patrignani:2016xqp}. We use the Higgs mass bounds \cite{Khachatryan:2016vau} within the range $m_{h}=[122,128] \mathrm{GeV}$ to account for the uncertainty of $m_h$ calculation in the MSSM, as stated in \cite{Allanach:2004rh}. Our low limit on the gluino mass is 2.2 TeV, which decays mostly into the third-generation squarks.  Also, we take the low mass bounds on the stop, sbottom, and first two-generation squarks as
1.25 TeV, 1.5 TeV, and 2 TeV, respectively.
We also consider constraints from rare decay processes: $B_{s}\rightarrow \mu^{+}\mu^{-}$ \cite{Aaij:2012nna}, $b\rightarrow s \gamma$ \cite{Amhis:2012bh}, and $B_{u}\rightarrow \tau\nu_{\tau}$ \cite{Asner:2010qj}. Moreover, the LSP neutralino relic density is within the $5\sigma$ range of the Planck collaboration \cite{Akrami:2018vks} measurement
In short, we choose
\begin{eqnarray}
&& m_{\tilde{g}}\geq 2.2~{\rm TeV}, m_{\widetilde t_1} \gtrsim ~ 1.25 \,{\rm TeV},  m_{\widetilde b_1} \gtrsim ~ 1.5 \,{\rm TeV} ~,~ \\
&& m_{\widetilde q} \gtrsim ~ 2 \,{\rm TeV}, m_h  = 122-128~{\rm GeV}~,~
\\
&& 0.114 \leq \Omega_{\rm CDM}h^2 (\rm Planck2018) \leq 0.126   \; (5\sigma)~,~
 \\
&& 0.15 \leq \frac{
 {\rm BR}(B_u\rightarrow\tau \nu_{\tau})_{\rm MSSM}}
 {{\rm BR}(B_u\rightarrow \tau \nu_{\tau})_{\rm SM}}
        \leq 2.41 \; (3\sigma)~,~
\\
&& 2.99 \times 10^{-4} \leq
  {\rm BR}(b \rightarrow s \gamma)
  \leq 3.87 \times 10^{-4} \; (2\sigma)~,~
\\
&& 0.8\times 10^{-9} \leq{\rm BR}(B_s \rightarrow \mu^+ \mu^-)
  \leq 6.2 \times10^{-9} \;(2\sigma)~.~
\end{eqnarray}

\begin{figure}[h!]
 \centering \includegraphics[width=7.90cm]{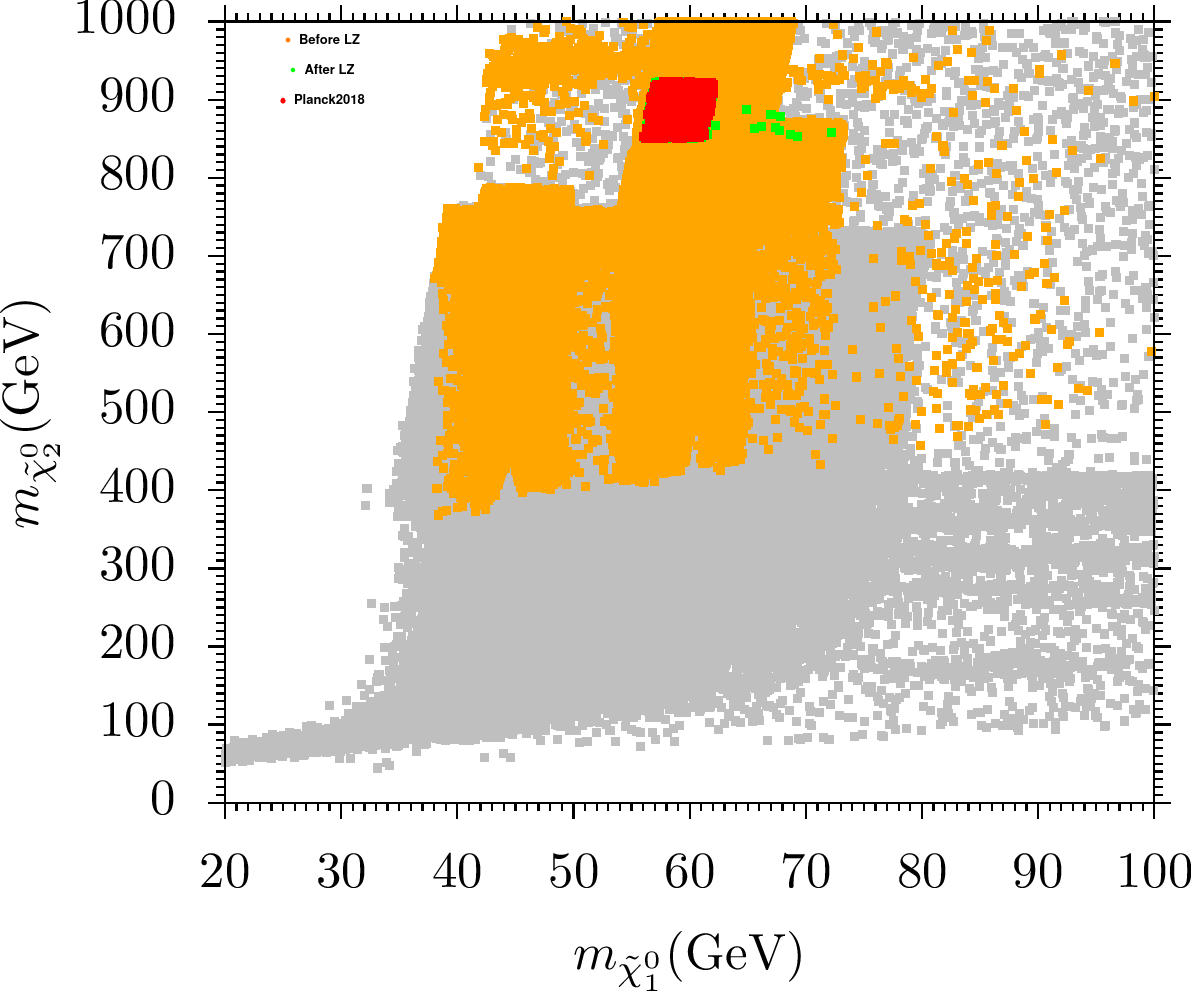}
    \centering \includegraphics[width=7.90cm]{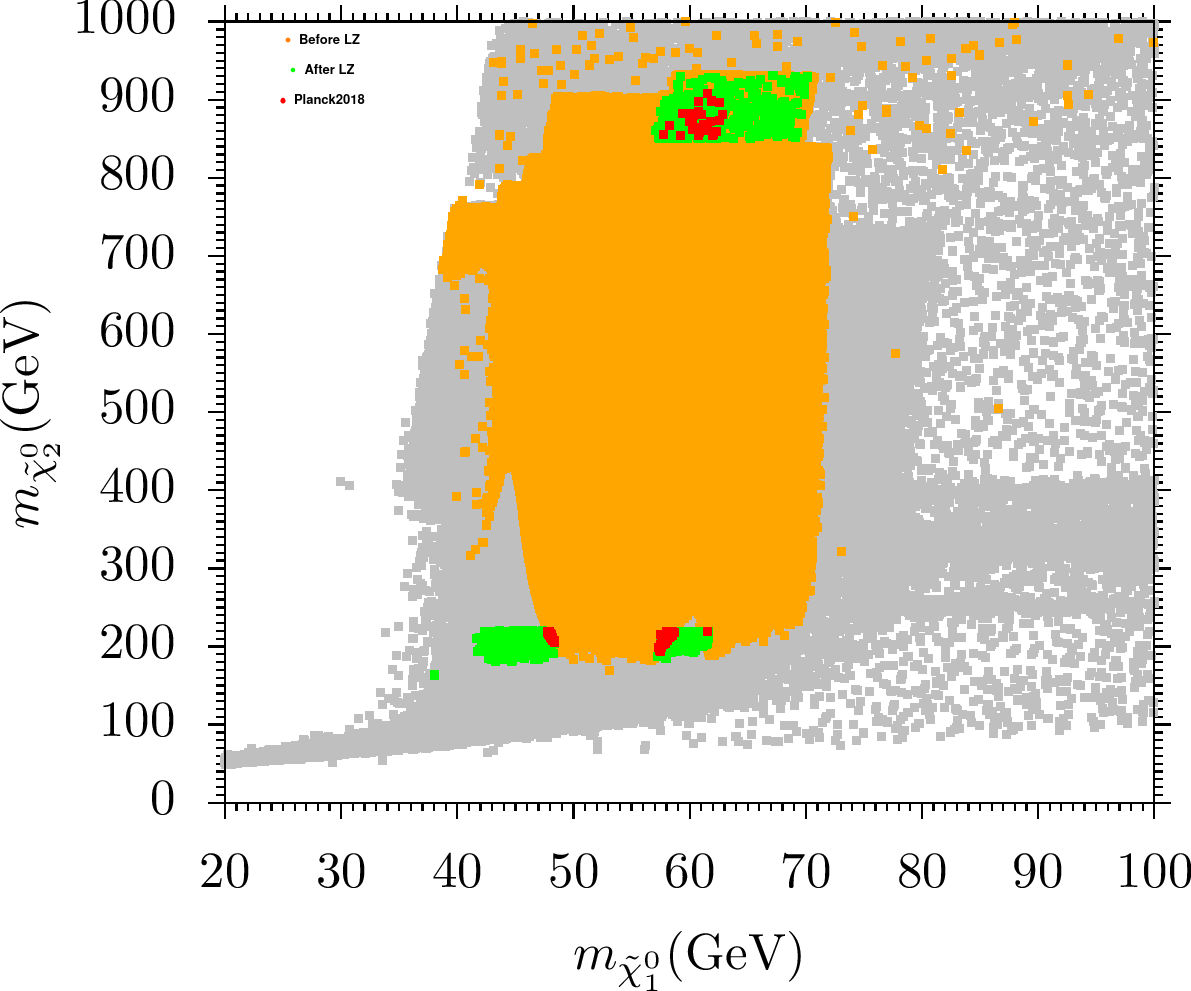}
	\caption{ We present $m_{\tilde{\chi}_1^0}$-$m_{\tilde{\chi}_2^0}$ plane for the positive $\mu$ (upper) and negative $\mu$ (lower). Grey points provide LSP neutralino and fulfill the REWSB. The orange points (\textbf{Before LZ}) represent the subset of gray points that satisfies the following constraints: sparticle LHC, Higgs, B-physics, LEP, oversaturated DM relic density bound, and DM DD from the PandaX-4T, XENON-1T, and PICO-60 experiments. DM DD constraints from the LZ experiment, electroweakino searches at the LHC, and undersaturated DM relic density bound are satisfied by green points and a subset of orange points (\textbf{After LZ}). Finally, the red points are a subset of green points and satisfy the saturated DM relic density bound (\textbf{Planck2018}).}
\label{F1}
\end{figure}

\begin{figure}[h!]
	\centering \includegraphics[width=8.90cm]{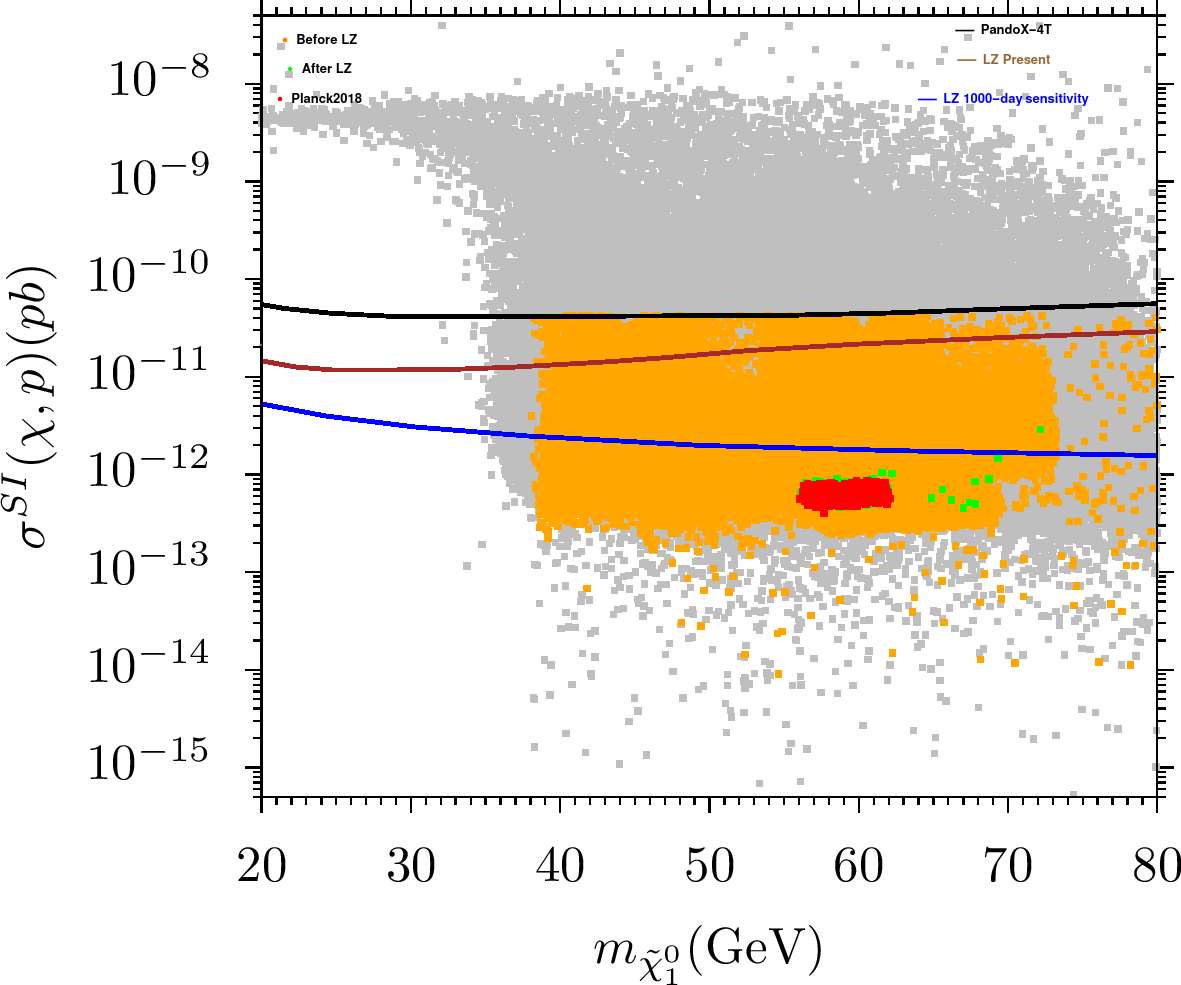}
	\caption{As a function of the LSP neutralino DM, the SD DM-nucleon cross-section for $\mu>0$ is presented. Grey points provide LSP neutralino and fulfill the REWSB. The orange points (\textbf{Before LZ}) represent the subset of gray points that satisfies the following constraints: sparticle LHC, Higgs, B-physics, LEP, oversaturated DM relic density bound, and DM DD from the PandaX-4T, XENON-1T, and PICO-60 experiments. DM DD constraints from the LZ experiment, electroweakino searches at the LHC, and undersaturated DM relic density bound are satisfied by green points and a subset of orange points (\textbf{After LZ}). Finally, the red points are a subset of green points and satisfy the saturated DM relic density bound (\textbf{Planck2018}).}
\label{F2}
\end{figure}
\begin{figure}[h!]
	\centering \includegraphics[width=8.90cm]{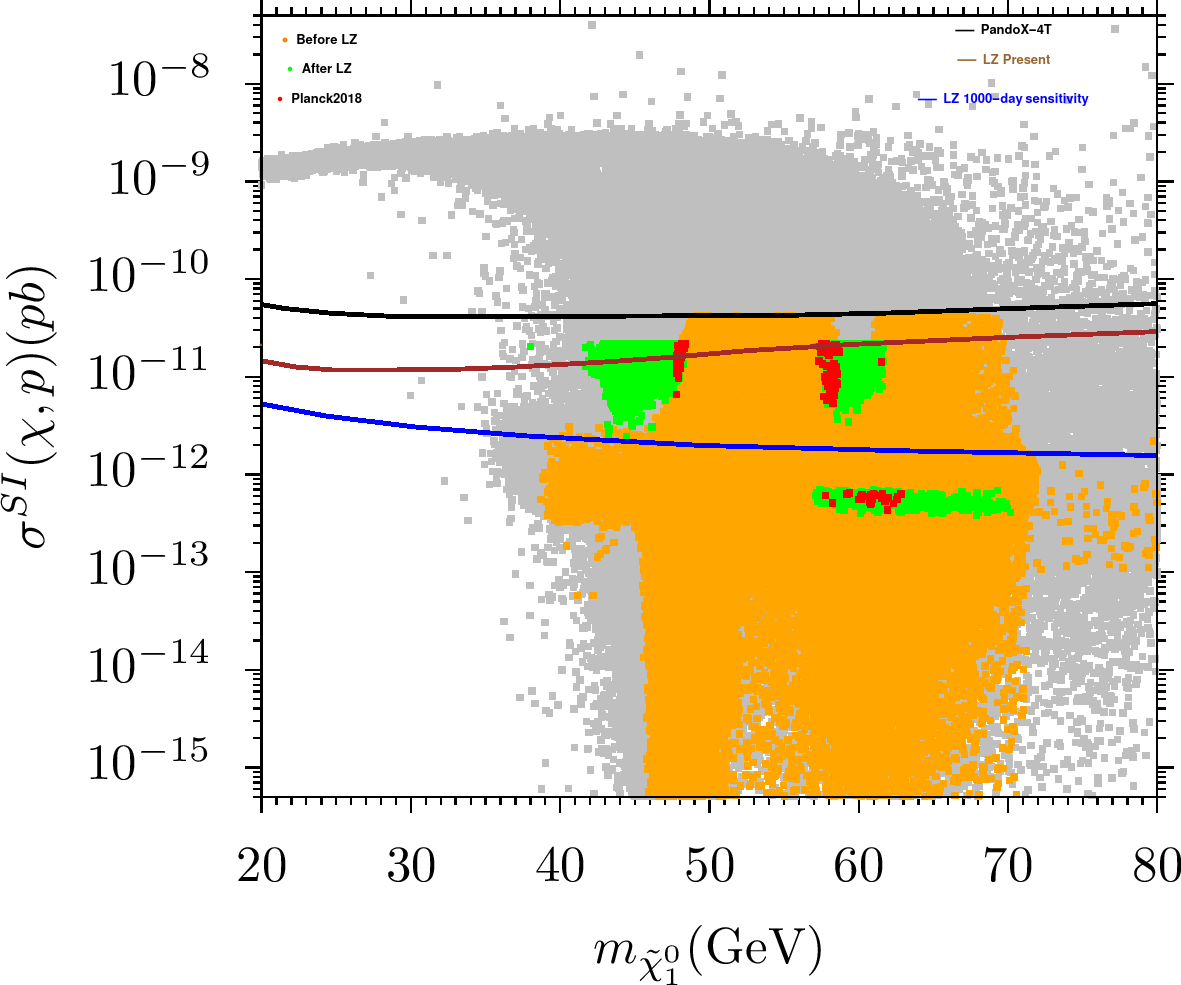}
        \centering \includegraphics[width=8.90cm]{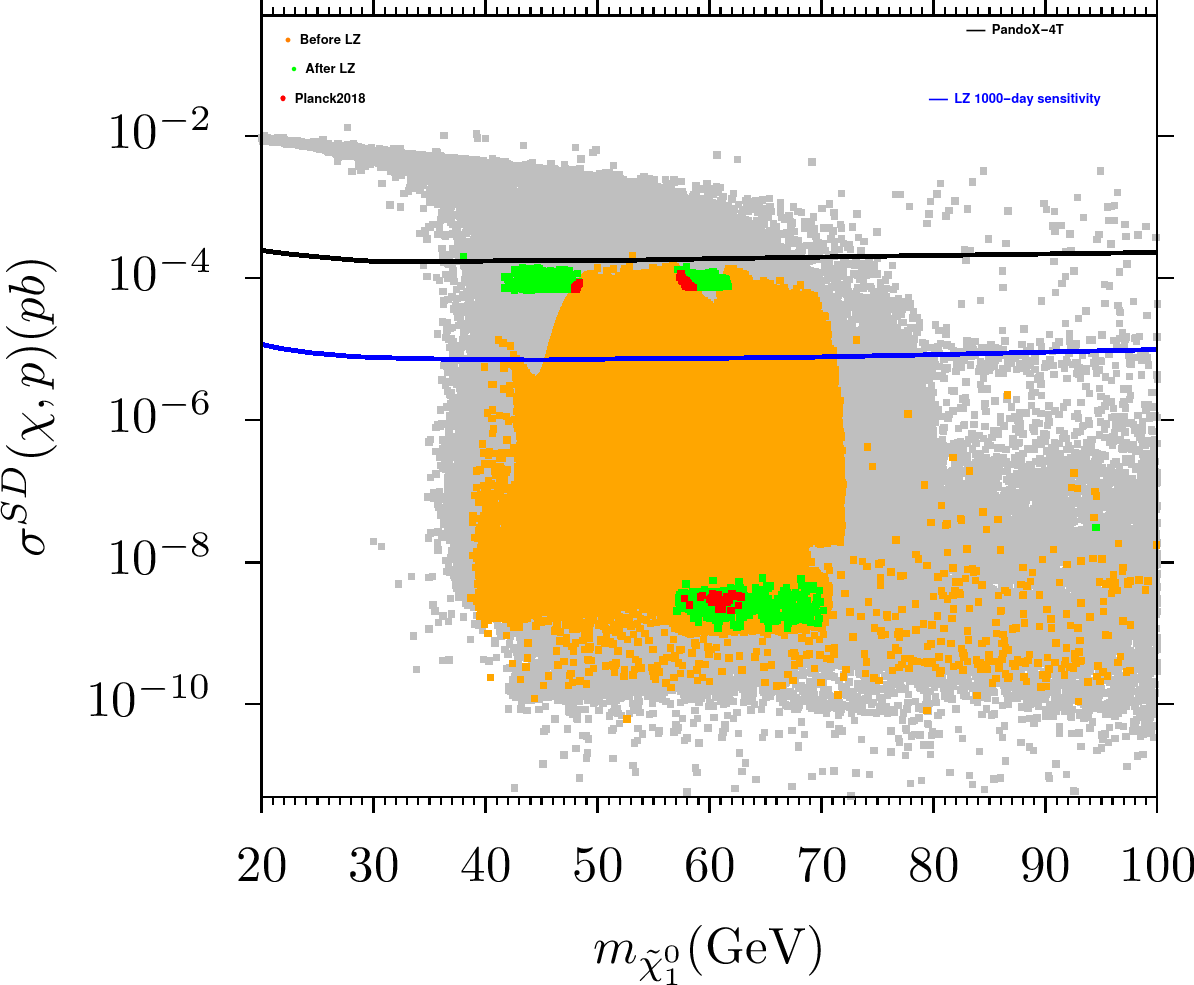}
	\caption{As a function of the LSP neutralino DM, the SI DM-nucleon cross-section (upper) and the SD DM-proton cross-section (lower) for $\mu<0$ are presented. Grey points provide LSP neutralino and fulfill the REWSB. The orange points (\textbf{Before LZ}) represent the subset of gray points that satisfies the following constraints: sparticle LHC, Higgs, B-physics, LEP, oversaturated DM relic density bound, and DM DD from the PandaX-4T, XENON-1T, and PICO-60 experiments. DM DD constraints from the LZ experiment, electroweakino searches at the LHC, and undersaturated DM relic density bound are satisfied by green points and a subset of orange points (\textbf{After LZ}). Finally, the red points are a subset of green points and satisfy the saturated DM relic density bound (\textbf{Planck2018}).}
\label{F3}
\end{figure}
\begin{figure}[h!]
        \centering \includegraphics[width=8.90cm]{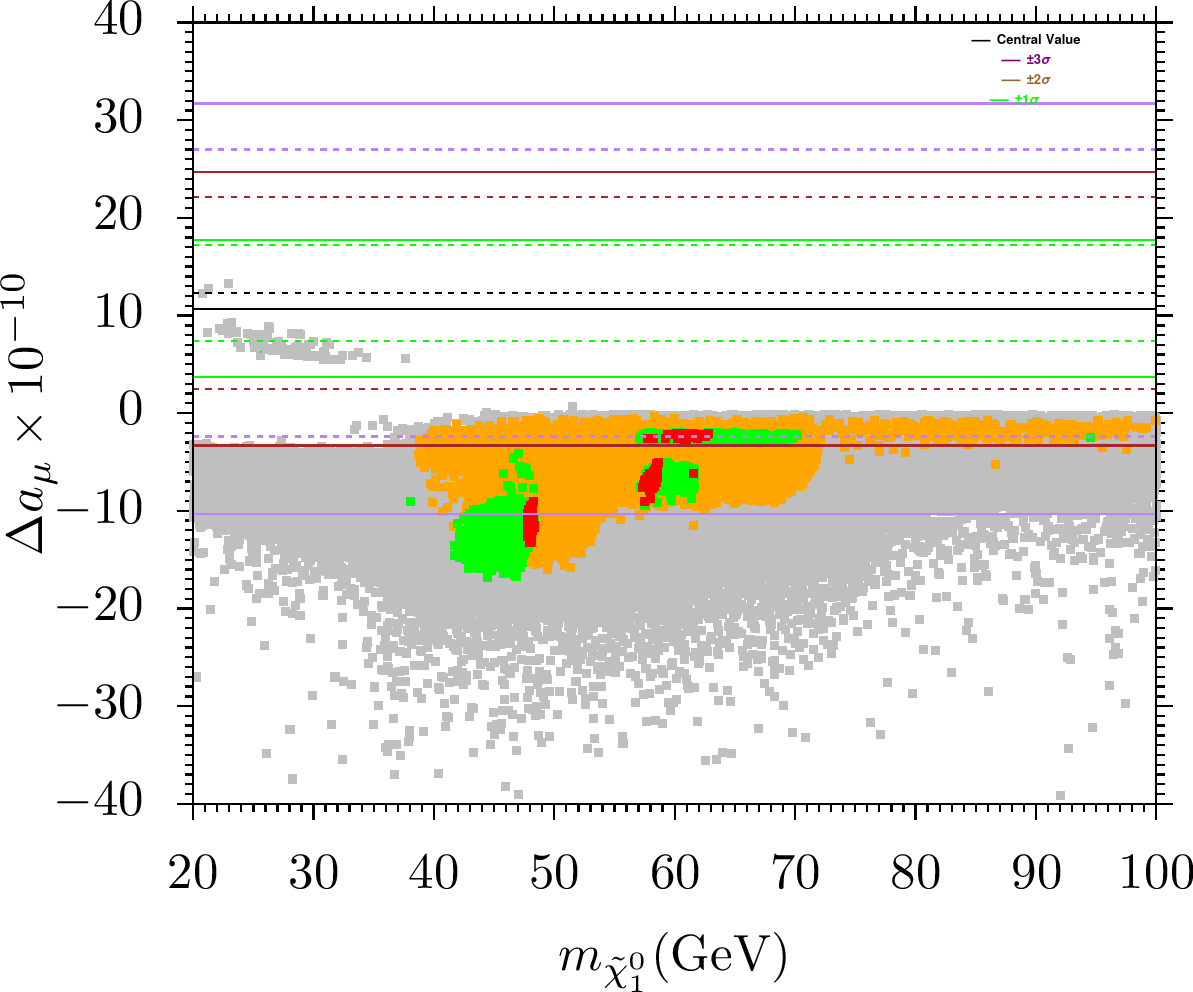}
	\caption{As a function of the LSP neutralino DM and $a_{\mu}$ are presented. Grey points provide LSP neutralino and fulfill the REWSB. The orange points (\textbf{Before LZ}) represent the subset of gray points that satisfies the following constraints: sparticle LHC, Higgs, B-physics, LEP, oversaturated DM relic density bound, and DM DD from the PandaX-4T, XENON-1T, and PICO-60 experiments. DM DD constraints from the LZ experiment, electroweakino searches at the LHC, and undersaturated DM relic density bound are satisfied by green points and a subset of orange points (\textbf{After LZ}). Finally, the red points are a subset of green points and satisfy the saturated DM relic density bound (\textbf{Planck2018}). The black line shows the central value of $\Delta a_{\mu}$ and the green, brown, and purple lines represent $1\sigma$, $2\sigma$, and $3\sigma$ deviation from the central value. The solid and dotted lines are for the BMW and CMD results, respectively.}
\label{F4}
\end{figure}
 \textbf{Results and Discussion:}
 We study the $H/Z$ resonances in the MSSM with the GmSUGRA for the higgsino mass parameter $\mu>0$ and $\mu<0$ scenarios. In addition to the constraints mentioned above, we also consider the constraints from ongoing direct detection (DD) DM experiments. These experiments impose constraints on the spin-dependent (SD) interactions between dark matter and neutrons (SDn), protons (SDp), and spin-independent (SI) DD cross-sections of the lightest neutralino LSP, which serves as the DM candidate. These cross-sections vary with the mass of the LSP. We use the ISAJET subroutine IsaTools \cite{Baer:2002fv} to calculate these cross-sections and subsequently compare them with experimental constraints from various DD DM experiments, including XENON-1T (SI \cite{XENON:2018voc}) and (SDp\cite{XENON:2019rxp}), PICO-60 ( SDp \cite{PICO:2019vsc}), PandaX-4T (SI \cite{PandaX-4T:2021bab}) and (SDp \cite{PandaX:2022xas}), and LZ (SI \cite{LZ:2022lsv}). Moreover, we take into account the findings from direct electroweakino searches conducted at the LHC. Recent investigations into electroweakinos, carried out by the CMS \cite{CMS:2020bfa} and ATLAS \cite{ATLAS:2021moa} Collaborations in leptonic final states, and by the ATLAS Collaboration \cite{ATLAS:2021yqv} in hadronic final states, have been instrumental in narrowing down potential mass ranges for $\tilde{\chi}_1^{\pm}$, $\tilde{\chi}_2^0$, and $\tilde{\chi}_3^0$. Notably, the ATLAS's analysis has extended sensitivity to higher mass regions, especially with the incorporation of hadronic final states. We apply constraints to explore our parameter space through a series of sequential steps. First, referred to as \textbf{“Before LZ" (orange points)}, we incorporate the constraints from various sources including LEP, the constraints from the Higgs measurements, B-physics bounds, oversaturated DM relic density bound from Planck2018, LHC  sparticle constraints, and direct detection experiments such as PandaX-4T, XENON-1T, and PICO-60, which is the subset of gray point. Subsequently, we include the constraint from the LZ experiment, electroweakino searches from the LHC, and undersaturated DM relic density bound from Planck2018 termed as \textbf{“After LZ" (green points)}, which is the subset of orange points. Finally, we integrate the saturated DM relic density constraints from Planck2018, denoted as \textbf{“Planck2018" (red points)}, which is the subset of green points.

In Fig. \ref{F1}, we depict the $m_{\tilde{\chi}_1^0}$-$m_{\tilde{\chi}_2^0}$ plane for both the $\mu>0$ and $\mu<0$ scenarios, after meeting the constraints labeled as \textbf{Before LZ} (orange), \textbf{After LZ} (green), and \textbf{Planck2018} (red). Our analysis reveals that the constraints \textbf{Before LZ} already exclude the solutions for the light $m_{\tilde{\chi}_2^0}\lesssim350$ GeV in the $\mu>0$ scenario. However, a region of parameter space exists in the $H-$funnel that remains viable despite all the constraints from the \textbf{Before LZ}, \textbf{After LZ}, and \textbf{Planck2018} for the heavy $m_{\tilde{\chi}_2^0}$. In this region, the spin-independent direct detection cross-section is slightly below the LZ 1000-day projection, representing a future experimental limit as depicted in Fig. \ref{F2}. We anticipate that the LZ experiment, with a 1000-day exposure, will probe this region in the near future. Fig. \ref{F1} highlights the impact of electroweakino searches, which impose a lower limit of $m_{\tilde{\chi}_2^0 }\gtrsim 850$ GeV for the $\mu>0$ scenario.

\begin{table}[h!]
	\centering
	\scalebox{0.6}{
		\begin{tabular}{|l|cccc|}
			\hline
			& Point 1 & Point 2 &Point 3& Point 4    \\
			\hline
			$m_{0}^{U}$          &   1977      & 1897&1891 & 1061  \\
			$M_{1},M_{2},M_{3} $         &   99.34,944.6,-1168.5      & 124.6, 962.7, -1132.6&127.7, 1024, -1216.8&121.6, 1020, -1226\\
			$m_{E^c},m_{L}$      &   787.2,166.3    &557.9, 143.4& 821.3, 155.3& 458.6, 118.5  \\
			$m_{H_{u}},m_{H_{d}}$           &    3253,2024     &  3342, 1382& 2002, 3672&2017, 2969 \\
			$m_{Q},m_{U^{c},m_{D^{c}}}$    & 1833.1,2552.3,2470 & 1746.6, 2449, 2406.3&1758.5, 2441.3, 2347.4&986.49, 1369.7, 1317.6 \\
			$A_{t}=A_{b},A_{\tau}$            &    4002,-442.4     & 4540, -365.9&4096, -422.7& 4382, -370.7 \\
			$\tan\beta$                      & 26.2 & 18.4 & 13.3 & 25.8\\
   $\mu$                      & -207,06 & -201.32 & -2060.2 & 1641.6\\
			\hline
			$m_h$            &  123    & 124&122&124   \\
			$m_H$            &  1552    & 1146&4183 & 2989 \\
			$m_{A} $         &  1542     &  1139& 4157& 2970   \\
			$m_{H^{\pm}}$    &  1555   & 1149 & 4184 & 2990    \\
			\hline
			$m_{\tilde{\chi}^0_{1,2}}$
			& 48,213 & 58,208& 63,881& 59, 870\\
			$m_{\tilde{\chi}^0_{3,4}}$
			& 220,816 &214,829& 2069,2069& 1643,1645 \\
			$m_{\tilde{\chi}^{\pm}_{1,2}}$
			&199,799  & 195,815&885,2074 &873,1648 \\
			\hline
			$m_{\tilde{g}}$  & 2640    &2560& 2739&2692 \\
			\hline $m_{ \tilde{u}_{L,R}}$
			& 2915,3313  & 2814,3230& 2964,3205& 2601,2625   \\
			$m_{\tilde{t}_{1,2}}$
			& 1853, 2109  & 1593,1981&2061,2428&1346,1872 \\
			\hline $m_{ \tilde{d}_{L,R}}$
			& 2916,3368 & 2815,3245&2965,3378&2602,2713\\
			$m_{\tilde{b}_{1,2}}$
			& 2053,3132 & 1939,3104&2399,3296&1855,2312 \\
			\hline
			$m_{\tilde{\nu}_{1}}$
			& 631       & 663&301&527  \\
			$m_{\tilde{\nu}_{3}}$
			& 544      & 636&68&244  \\
			\hline
			$m_{ \tilde{e}_{L,R}}$
			& 640,742   & 673, 423&337,1153&245,713 \\
			$m_{\tilde{\tau}_{1,2}}$
			& 557,605    & 365, 649&161,1093&112,410\\
			\hline
			$\Delta a_{\mu}$
			& -1$\times 10^{-9}$ & -6.1$\times 10^{-10}$ &-2.1$\times 10^{-10}$&4.2$\times 10^{-10}$  
			\\
			\hline
			$\sigma_{SI}(pb)$
			& 1.3$\times 10^{-11}$ & 1.6$\times 10^{-11}$&6.3$\times 10^{-13}$&7.7$\times 10^{-13}$\\
   $\sigma_{SD}(pb)$
			& 7.2$\times 10^{-5}$ & 8.5$\times 10^{-5}$ &3.2$\times 10^{-9}$&8.9$\times 10^{-9}$ 
			\\
			$\Omega_{CDM}h^2$
			& 0.115      & 0.123&0.115&0.124 \\
			\hline
			\hline
		\end{tabular}
  }
 \caption{The table provides the masses in the unit GeV}
		\label{table1}
\end{table}

Let's shed light on the $\mu<0$ scenario. As depicted in Fig. \ref{F1}, there is a significant difference between the scenarios with positive $\mu$ and negative $\mu$. The \textbf{Before LZ} constraint excludes the presence of lighter higgsinos (with masses $\lesssim$ 350 GeV) for $\mu>0$, while they remain viable in the $\mu<0$ scenario. These lighter higgsinos manage to survive while adhering to all current constraints in the $\mu<0$ scenario for both $H$ and $Z$ poles. This is because in the MSSM, for $\mu<0$, the SI DD cross-section is reduced due to a cancellation between the contributions of the two CP-even neutral Higgs bosons ($h$ and $H$) in diagrams involving down-type quarks \cite{Ellis:2000ds,Baer:2003jb}. In contrast, for $\mu>0$, these contributions demonstrate constructive interference, leading to an enhancement in the SI DD scattering cross-sections. A narrow parameter space permitted by the LZ experiment and the electroweakino searches is observed in the $H$ and $Z$ poles regions for the negative $\mu$ case, specifically for the lighter $m_{\tilde{\chi}_2^0}$. While green points solutions appear predominantly for heavy higgsinos in both $\mu>0$ and $\mu<0$ scenarios, the survival of such solutions for the lighter $m_{\tilde{\chi}_2^0}$ is evident only in the $\mu<0$ scenario. However, adherence to the relic density constraint (\textbf{Planck2018}) results in red solutions around $m_{\tilde{\chi}_1^0} \sim 45$ GeV and $m_{\tilde{\chi}_1^0} \sim 60$ GeV. These solutions correspond to well-known Z pole and Higgs pole scenarios, where two LSP neutralinos annihilate via s-channel exchange of a virtual particle, such as a Z or Higgs boson, with the mass of the exchanged particle approximately twice the LSP neutralino mass. It is evident from the recent LZ findings and electroweakino searches that the majority of the $H$ and $Z$ poles regions have been excluded for the light higgsinos, leaving behind only a small permissible area where the $m_{\tilde{\chi}_2^0}$ is either very small (approximately $175-215, {\rm GeV}$) or greater than 850 GeV, as depicted in the lower plot of Fig. \ref{F1}. Furthermore, the entire $Z$ pole region lies comfortably within the projected sensitivity of the current LZ experiment, as presented in the upper plot of Fig. \ref{F3}. Note, we want to comment here that we keep the LZ present bound a bit relaxed to see the picture more clearly in this region, as our $Z$ pole is severely constraint by the LZ present bound for the solutions to satisfy the \textbf{Planck2018}  constraint (red solutions). On the contrary, the $H$ pole region remains relatively unaffected by the LZ constraints, with a significant portion residing comfortably below the present reach of LZ for light higgsinos. In fact, with its full 1000-day sensitivity, the LZ experiment is expected to explore our parameter space containing light higgsino in the $H$ and $Z$ poles solutions. Interestingly, the SD-DD cross-section is directly proportional to the Z boson and Lightest Supersymmetric Particle (LSP) coupling square, which depends on the higgsino component of $\tilde{\chi}_1^0$. Consequently, the lighter higgsinos exhibit smaller values of the SD DM cross-section compared to their heavier counterparts, leading to the emergence of two distinct red solutions in the $H$ pole region in the lower plot of Fig. \ref{F3}. It is important to note that detecting a DM signal in the $Z$ pole from DD experiments would probably suggest $\mu$ negative while for the $H$ pole, this is the case only for the light higgsino. To ascertain the sign of $\mu$, additional signal detection in collider experiments would be necessary to validate a DM signal in the $H$ pole for the heavy higgsino. To be more precise, light higgsinos would provide evidence that $\mu<0$, whereas heavy higgsinos would leave this ambiguity. There has been a long-standing discrepancy between the theoretical prediction and the experimental measurement of the muon’s anomalous magnetic moment \( a_\mu \equiv (g - 2)_\mu/2 \). The most current experimental global average, derived from BNL and Fermilab (BNL+FNAL), is \cite{Muong-2:2023cdq,Muong-2:2021ojo}
\[
a_{\text{BNL+FNAL}}^\mu = 116592059(22) \times 10^{-11},
\]
while the data-driven theoretical calculation yields
\[
a_{\text{SM}}^\mu = 116591810(43) \times 10^{-11}.
\]
By merging the data-driven theoretical calculation with the experimental results, the observed discrepancy is
\[
\Delta a_{\text{BNL+FNAL}}^\mu = (24.9 \pm 4.8) \times 10^{-10}.
\]
Despite the widely recognized data-driven value, recent lattice calculations have shown a smaller difference of $\Delta a_\mu = (10.7 \pm 6.9) \times 10^{-10}$ \cite{Borsanyi:2020mff,Kuberski:2023qgx}. The discrepancy reported by CMD-3, which amounts to $\Delta a_\mu = (4.9 \pm 5.5) \times 10^{-10}$ \cite{CMD-3:2023rfe} 
 is based on the \( e^+ e^- \to \pi^+ \pi^- \) channel, which motivated recalculations of previous data, giving $
\Delta a_\mu = (12.3 \pm 4.9) \times 10^{-10}$ \cite{Davier:2023fpl}. The sign of the Higgsino mass parameter \( \mu \) plays a pivotal role in SUSY GUTs, especially when considering its impact on the muon anomalous magnetic moment \( a_\mu \equiv (g - 2)_\mu/{2} \). As shown in Fig.~\ref{F4}, e present the results from an extensive scan across the parameter space, providing insight into the observed deviation \( \Delta a_\mu \). Our findings indicate that the SUSY contribution from GmSUGRA is consistent with the central value of \( (g-2)_\mu \) with deviations up to \( 2\sigma \) for the BMW results and up to and \( 3\sigma \) for the CMD experimental data \cite{Borsanyi:2020mff,Kuberski:2023qgx,CMD-3:2023rfe,Davier:2023fpl}. Finally, we provide four benchmark points in table-\ref{table1}, illustrating our findings. Point 1 and Point 2 correspond to solutions with light $\tilde{\chi}_2^0$ for the $\mu<0$ case in the $Z$ and $H$ poles respectively. Point 3 represents a heavy $\tilde{\chi}_2^0$ solution in the $H$ pole consistent with our existing constraints. Finally, Point 4 depicts the heavy $\tilde{\chi}_2^0$ solution for the $\mu>0$ scenario.

\textbf{Conclusion:}
In summary, this letter highlights how recent experiments, especially the findings from DM DD measurements at LZ and LHC electroweakino searches, have significantly restricted the scenario where $\mu>0$. Only very heavy higgsinos are permitted in the $Z$ and $H$ poles, while the $Z$ and $H$ poles are entirely ruled out for light $m_{\tilde{\chi}_2^0}$. In the $\mu<0$ case for a light neutralino thermal DM from the GmSUGRA in the MSSM, the allowable parameter space is constrained to either a narrow region of light higgsinos with masses ranging from $175$ to $215$ GeV in both the $H$ and $Z$ poles, or to higgsinos heavier than approximately $850$ GeV. The present state of light higgsinos from the current electroweakino restrictions remains uncertain in the mass range of $175-215$ GeV. Thus, experimental cooperation in this area is essential to provide conclusive answers about the existence of light higgsinos, taking into account all possible decay modes, including $WZ$. This offers the LHC Run-3 a compelling target. Our results also exhibit sensitivity to the ongoing tension in the muon anomalous magnetic moment, \( (g-2)_\mu \), underscoring the contribution of SUSY to the observed discrepancy between experimental measurements and Standard Model predictions. In conclusion, we find ourselves at a really intriguing crossroads: either the hypothesis of a light thermal neutralino DM in the MSSM is utterly ruled out by the planned tests in the near future, or we are very close to witnessing the first direct signs of novel physics at the LHC.

\textbf{Acknowledgments:--} TL is supported in part by the National Key Research and Development Program of China Grant No. 2020YFC2201504, by the Projects No. 11875062, No. 11947302, No. 12047503, and No. 12275333 supported by the National Natural Science Foundation of China, by the Key Research Program of the Chinese Academy of Sciences, Grant NO. XDPB15, by the Scientific Instrument Developing Project of the Chinese Academy of Sciences, Grant No. YJKYYQ20190049, and by the International Partnership Program of Chinese Academy of Sciences for Grand Challenges, Grant No. 112311KYSB20210012.


\end{document}